# Multidimensional Analysis of System Logs in Large-scale Cluster Systems

Wei Zhou, Jianfeng Zhan, Dan Meng
*National Research Center for Intelligent Computing Systems,
Institute of Computing Technology, Chinese Academy of Sciences
Email: {zhouwei, jfzhan, md}@ncic.ac.cn*

**Abstract**

*It is effective to improve the reliability and availability of large-scale cluster systems through the analysis of failures. Existed failure analysis methods understand and analyze failures from one or few dimension. The analysis results are partial and with less precision because of the limitation of data source. This paper presents multidimensional analysis based on graph mining to analyze multi-source system logs, which is a promising failure analysis method to get more complete and precise failure knowledge.*

## 1. Introduction

As the growing need of large-scale cluster systems in scientific computing and commercial applications, errors and failures become normal. They derive from software, hardware maintenance, operations, environment, and process and so on [1].

We can get the rules and the patterns of errors and failures by analyzing the system logs. The results which we called failure knowledge can be used to understand the failure behavior, detect faults and identify failures quickly, predict failures in order to minimize their worst impact, diagnose and respond to failures effectively, model and improve RAS (Reliability, Availability, and Serviceability) metrics.

There existed some log analysis methods that can be reduced to four categories. Statistics-based methods use statistical methods to find the laws of failure, which can be used to predict failures [1][2]. Model-based methods use data mining combining artificial intelligence to construct a probabilistic or analytical model of the system. A warning is triggered when a deviation from the model is detected [3][4]. Relation-based approaches correlate explicit events such as traces and faults, attributes as SLO with implicit user behavior and status [5][6][7][8]. Path-based approaches correlate applications between nodes by analyzing communication paths [9][10].

This research is supported by the National Natural Science Foundation for Young Scientists of China (Grant No. 60703020)

All these current log analysis methods get failure knowledge from only one or few aspect such as time series, spatial series, communication, events, errors and operations. Because of the limitation of data source, the analysis results of these analysis methods are partial and with less precision.

In this paper we present a new approach to analyze multi-source system logs. We use weight-based graph mining technology to get the detailed relationships of multidimensional data. The more complete and precise failure knowledge can improve the accuracy of the applications of failure knowledge.

The rest of this paper is organized as follows. Section 2 introduces multidimensional Analysis. We present a novel framework to analyze system logs in section 3 and describe the failure knowledge mining based on graph mining in section 4. The paper concludes with Section 5.

## 2. Multidimensional Analysis

Failure analysis requires multiple perspectives on the data, such as events, failures, and status. Each of these perspectives is called dimension. Multidimensional analysis analyzes multi-source system logs. It can get more complete perspective of the failures and more precise failure knowledge.

The system logs collected by some tools can be summed up to several dimensions. The data sources include events of systems and applications, statuses of systems and applications, communications of systems and applications, RAS performance of applications.

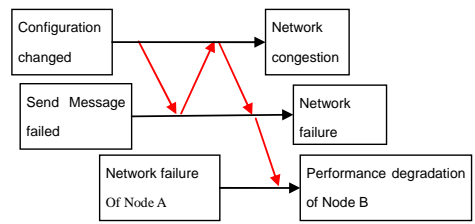

Figure 1. Example of Multidimensional Analysis

For example, we can use current log analysis methods to get three failure patterns showed as the black lines in Figure 1. Multidimensional analysis can find the relationships of complex multidimensional data. The analysis result is showed as the red arrow lines in Figure 1. From the figure we can find that the configuration change of Node A is a possible root cause of performance degradation of Node B.

## 3. Framework

The traditional log analysis methods have four steps: (1) Log collection. All system logs are summed up to some dimensions. (2) Data preprocessing. In this step, we can filter system logs by removing repeated data, clearing the noises, and so on, and get simple operational context to denote complex system logs. (3) Failure analysis. The log analysis methods are presented in Section 1. (4) Applications. The analysis results can be used on failure detection, failure prediction, fault diagnosis, RAS quantification.

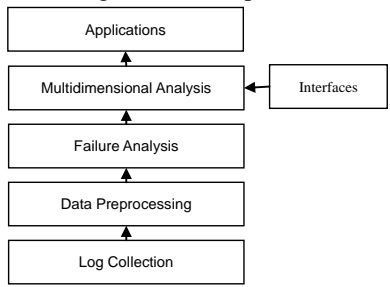

Figure 2. Framework of Multidimensional Analysis

The new framework adds two parts:

(1) Multidimensional analysis. Using graph mining technologies [11], we can get useful and precise knowledge from more dimensions.

(2) Interfaces. We can integrate failure rules and patterns produced by others log analysis methods, or some expert knowledge, and get more precise failure knowledge.

## 4. Failure Knowledge Mining

Graph mining is the extraction of novel and useful knowledge from a graph representation of data [11]. Graph mining can deal with data from a single, time-varying real number to a complex interconnection of entities and relationships. Graph mining can be used to mine relationships of failures from multidimensional data sources.

The research includes these key points:

(1) Failure analysis deals with each dimension at first, and get a few of rules. The rules denote the frequent time series of the dimension. Each rule has support and confidence.

(2) Time processing. The rules in one dimension are order, but not order among the dimensions. We should sort the rules to time series by traversing the data the second time.

(3) Failure knowledge mining. In this step, we use weight–based frequent subgraph mining to get the failure knowledge from the dimensions. The weight here means support and confidence of the rules.

(4) Calculation of support and confidence. We calculate support and confidence of failure knowledge through that of the one-dimension rules.

(5) The integration. We can integrate other analysis results and expert knowledge into the rules.

## 5. Conclusion

Errors and failures often happen in large-scale cluster systems. We can improve the reliability and availability of the systems effectively through the analysis of failures.

At present the failure analysis methods analyze the data from on or few dimension. The results are partial and with less precision. Multidimensional analysis based on graph mining can analyze the data from multiple dimensions, integrate other analysis methods and expert knowledge, and get more complete and precise failure knowledge. Failure knowledge can be used to predict and detect and diagnose failure, quantify RAS performance of Internet Services, and achieve high reliability and availability of the systems.

Future work focuses on the research, implement, and evolution of weight-based graph mining algorithm to achieve precise failure knowledge.